\documentclass[aps,prb,floatfix,twocolumn,amssymb,nobibnotes,longbibliography,nofootinbib]{revtex4-1}
\usepackage{graphicx}
\usepackage{float}
\usepackage{amsmath}
\usepackage{natbib}
\begin{document}
\title{Quantum critical points, lines and surfaces}
\author{Hui Yu and Sudip Chakravarty}%
\affiliation{Mani L.Bhaumik Institute for Theoretical Physics \\Department of Physics and Astronomy, University of California Los Angeles, Los Angeles, California 90095, USA}
\date{\today}%


\begin{abstract}
In this paper we promote the idea of quantum critical lines ({\em inter alia} surfaces) as opposed to points. A quantum critical line obtains when criticality at zero temperature is extended  over a continuum in   a one-dimensional line. We base our ideas on a simple but exactly solved model introduced by one of the authors involving a one-dimensional quantum transverse field Ising model with added 3-spin interaction. While many of the ideas are quite general, there are other aspects that are not. In particular, a line of criticality with continuously varying exponents is not captured. However, the exact solvability of the model gives us considerable confidence in our results. Although the pure system is analytically exactly solved, the disorder case requires numerical analyses based on exact computation of the correlation function in the Pfaffian representation. The disorder case  leads to dynamic structure factor as a function of frequency and wave vector. We expect that the model is experientally realizable and perhaps many other similar models will be found to explore quantum critical lines.
\end{abstract}
\maketitle

\section{INTRODUCTION}
 Quantum critical point (QCP)\cite{Hertz:1976} is an important widely discussed topic.\cite{Continentino:2017,Sachdev:2005} Yet there are few exact solutions that we can rely on. In their absence, it is difficult to  assert with certainty the properties that are implied, especially in regard to experimental systems.  However, there are some limitations of this concept:  (a) fine tuning to the quantum critical point, a related issue is the extent of the quantum critical fan;  (b) the range of temperature in which the concept should hold. After all, experiments are carried out at finite temperatures, and the quantum behavior in the ground state must be inferred from these measurements. QCP does afford  views of quantum fluctuations at zero temperature from measurements at finite temperature. This idea was  effectlvely exploited~\cite{Chakravarty:1989,Chakravarty:1988} in the problem of two dimensional quantum antiferromagnets, and  it proved that  the ground state has  long range antiferromagnetic order.~\cite{Birgeneau:1988}
 
\begin{figure}[htb]
\includegraphics[scale=0.7]{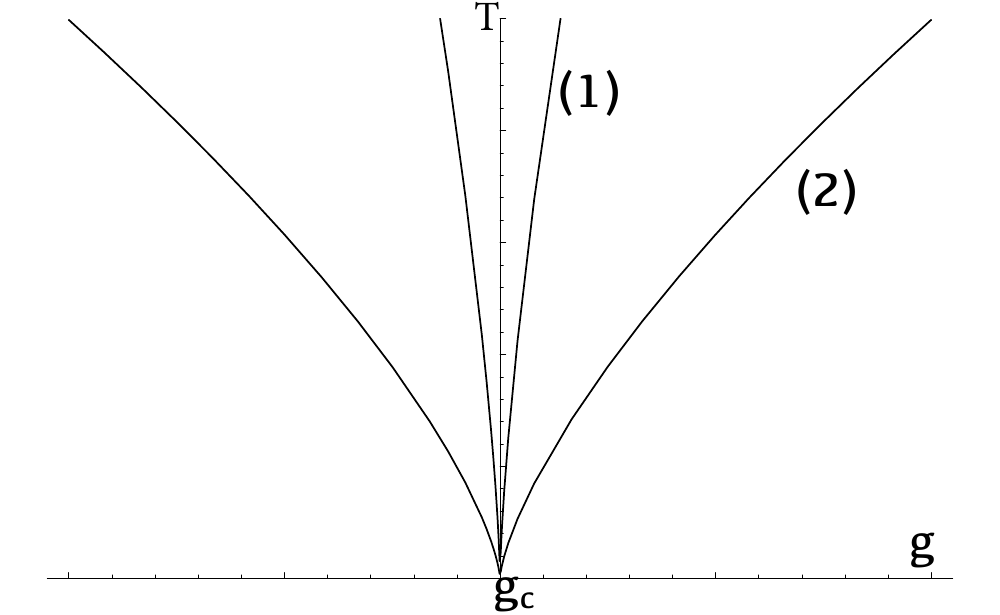}
\caption{A  sketch of a quantum critical point and  associated quantum critical fans.  $T$ is the temperature and $g$ is the tuning parameter. (1) and (2) are two possible quantum critical fans corresponding to two different cusp exponents. $g_c$ is the quantum critical point. The figure simply illustrates how the width of the quantum critical fans restrict the fine tuning.}.
\label{fig:QCP}
\end{figure}

The issue  regarding  fine tuning  is shown {\em  schematically} in  Fig.~\ref{fig:QCP}. The cusp, introduced in Refs. \onlinecite{Chakravarty:1989,Chakravarty:1988}, specifies the criticality of  the quantum Hesienberg model in $(2+1)$ dimensions; the extra dimension beyond 2 spatial dimensions corresponds to the tmporal dimension reflecting the quantum nature of the model. The exponent of the cusp, $\nu_{d+1}$,  is  difficult to establish in general and  therefore the extent of the quantum critical fan; a cusp is not the only possibility and depends on the crossover exponent. A finite temperature experiment may not be able to detect the quantum criticality because of the limitation of fine tuning. 

\begin{figure}[h]
\includegraphics[scale=0.35]{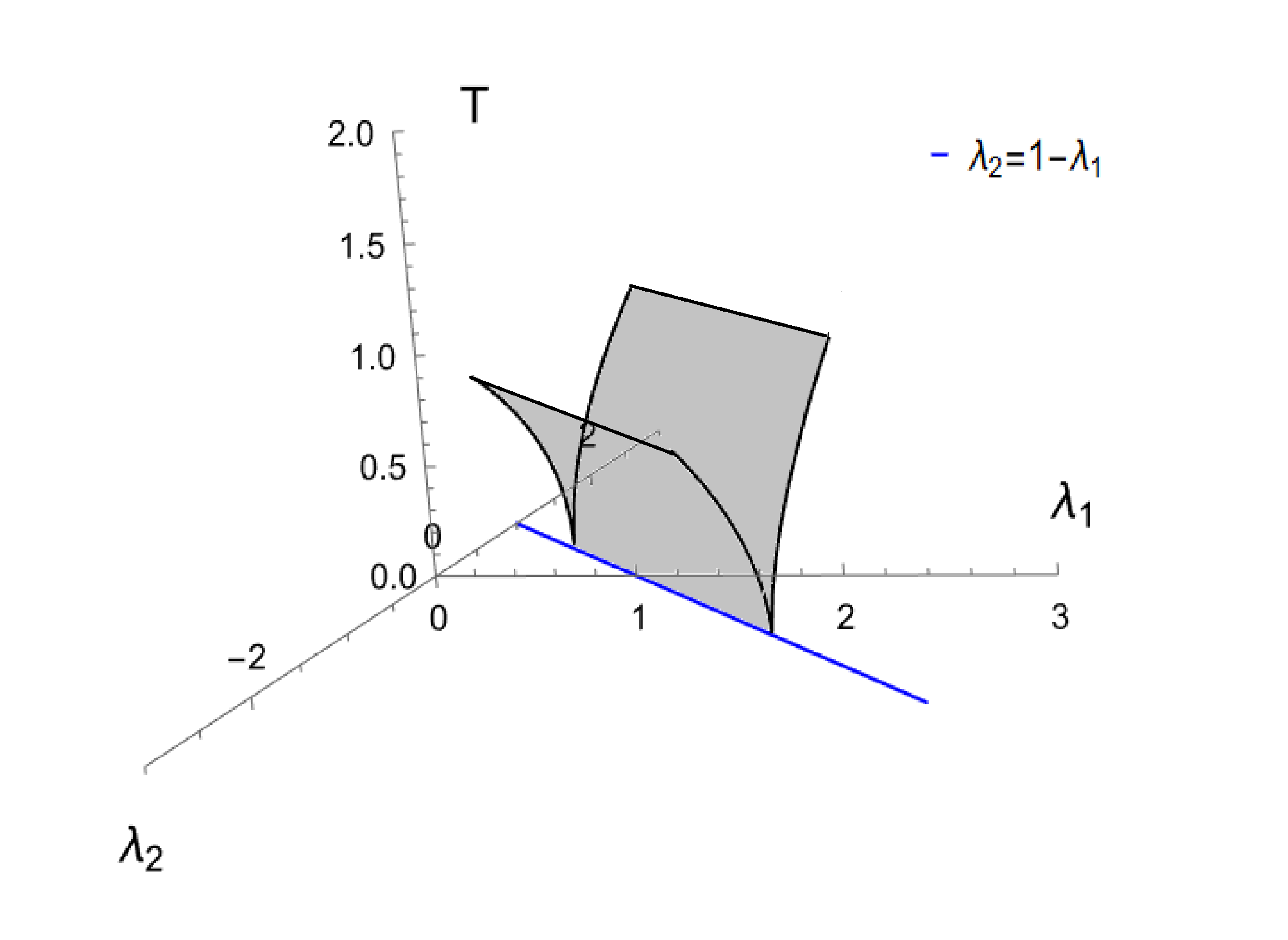}
\caption{A  sketch of a quantum critical line $\lambda_2=1-\lambda_1$, extended to finite temperature $T$.}.
\label{QCL-T}
\label{fig:QCL}
\end{figure}

On the other hand if the criticality were stretched out on a line at zero temperature, as shown in Fig.~\ref{fig:QCL} and not confined to a point, experimental signature of quantum fluctuations could  be more readily  observed at finite temperatures, and this is what we want to emphasize.

There is a classic example of QCP, known as the transverse field Ising model  in one dimension whose exact solution provides  the properties of a quantum critical point addressed in detail in Ref.~\onlinecite{Sachdev:2005}. Experimentally, this model is well represented by 
$\mathrm{CoNb_{2}O_{6}}$,\cite{Kinross:2014} which illustrates the nature of quantum criticality. Quantum critical lines (QCL) or surfaces (QCS) are less studied. We introduce a simple but {\em exactly  solved} model of QCL and explore it in some detail. This  model  was  introduced by Kopp and Chakravrty,~\cite{Kopp:2005} and has numerous exciting properties. It  remains to  be experimentally studied. The phase transitions in this model  have  intriguing topogical aspects discussed by Niu {\em et al.}~\cite{Niu:2012} to which we shall  return in the text. We cannot overemphasaize the fact that  subtle effects related to quantum fluctuations require an exact solution of the proposed model. It is hoped that an experimental realization will be found and studied. At the end we briefly remark on the Fermi surface as a QCS. The discussion is based on  non-interacting fermions that reflect criticality because of the infinitely sharp nature of the fermi surface.

The  paper is organized as follows: In Section II, we introduce the model and its quantum critical lines, and explore its properties in the context of QCL. In the following Section III we briefly touch upon Majorana representation, which was extensively discussed in the past~\cite{Niu:2012} and in Sec. IV the effect of disorder is treated, as any experimental realization will evidently involve randomness of model parameters.  In Sec. V we briefly touch upon Fermi surface from the present perspective of criticality. The final Section VI is a summary.

\section{The Model and its quantum critical lines}

The model we study is the three spin extension of the TFIM, studied previously by~\cite{Kopp:2005}. We will focus on topics that was not studied previously. The Hamiltonian, $H$, is
\begin{equation}
H=-\sum_i(h_i \sigma_i^x+\lambda_2\sigma_i^{x}\sigma_{i-1}^z\sigma_{i+1}^z+\lambda_1\sigma_i^z\sigma_{i-1}^z)
\end{equation}
written in terms of  standard Pauli matrices.  
In this section we discuss its phase diagram.  We shall set $h_i= h= cst.$ The Hamiltonian 
 after Jordan-Wigner~\cite{Pfeuty:1970},\cite{Lieb:1961}
transformation 
\begin{eqnarray}
\sigma_i^x = 1-2c_i^\dagger c_i \\
\sigma_i^z=-\prod_{j<i}(1-2c_j^\dagger c_j)(c_i +c_i^\dagger)
\label{Eq:JW-2}
\end{eqnarray}
is
\begin{equation}
\label{eq:Ham1}
\begin{split}
H&=-\sum_{i=1}^{N}h (1-2c_i^\dagger
c_i)-\lambda_1\sum_{i=1}^{N-1}(c_i^\dagger c_{i+1}+c_i^\dagger
c_{i+1}^\dagger+h.c.)\\&-\lambda_2\sum_{i=2}^{N-1}(c_{i-1}^\dagger
c_{i+1}+c_{i+1}c_{i-1}+h.c.).
\end{split}
\end{equation}
In contrast to the spin model, the spinless fermion Hamiltonian is actually a one-dimensional {\em mean field} model\cite{Kitaev:2001} of a $p$-wave  superconductor, but there are both nearest and next nearest neighbor hopping,
as well as condensates---note the pair creation and destruction  operators.  The solution of the corresponding spin Hamiltonian through Jordan-Wigner transformation  is, however, exact and includes all possible fluctuation effects and is {\it not a mean field solution of any kind}.

Imposing periodic boundary conditions, a Bogoliubov transformation results in its diagonalized form:
\begin{equation}
H=\sum_{k} 2E_{k}\left(\eta_{k}^{\dagger}\eta_{k}-1\right).
\end{equation}
As usual, the anticommuting fermion operators $\eta_{k}$'s are suitable linear combinations in the momentum space of the original Jordan-Wigner fermion operators. The spectra of excitations are (lattice spacing will be set to unity  throughout the paper unless  stated otherwise)
\begin{equation}
E_{k}=\sqrt{1+\lambda_{1}^{2}+\lambda_{2}^{2}+2\lambda_{1}(1-\lambda_{2})\cos k -2\lambda_{2}\cos 2k}
\end{equation}
 We haveset $h=1$ when discussing the zero temperature, $T=0$, properties.. At finite temperaures we must keep $h$ to be finite, similarly for the disordered system (cf. below).
Quantum phase transitions of this model are given by the nonanalyticities of the ground state energy:
\begin{equation}
E_{0}=-2\sum_{k}E_{k}. 
\end{equation}
Taking the derivative of the energy dispersion, we get:
\begin{equation}\frac {\partial E_k}{\partial k}= \frac{4 \text{$\lambda_2 $}
   \sin 2 k-2 \text{$\lambda_1
   $} (1-\text{$\lambda_2 $})
   \sin k}{2 \sqrt{2
   \text{$\lambda _1$}
   (1-\text{$\lambda_2 $}) \cos
   k-2 \text{$\lambda_2 $}
   \cos 2 k+\text{$\lambda_1
   $}^2+\text{$\lambda_2
   $}^2+1}}
   \end{equation}
 The derivative vanishes   vanishes at $k=0,\pm\pi$ and $\cos k= \lambda_1(1-\lambda_2)/4 \lambda_2$. For these values of $k$  the spectra  assume minimum values. 
The nonanalyticites are  defined by the {\em critical lines} where the gaps collapse and the $T=0$ correlation length diverges; see Fig.~\ref{fig:fig1}. Note that the change in the number of Majorana zero modes at the end of the chain corresponds to  topological phase transitions.

To follow the description below refer to Fig.~\ref{fig:fig2}.
The contour plot of the square of the gap for $0\le \lambda_1\le 2$ and $0\ge \lambda_2 \ge -1$ is given in Fig.~\ref{fig:Contour}.
We enumerate below some of the features of the phase diagram and its associated critical lines:
\begin{figure}[htbp]
\begin{center}
\includegraphics[scale=0.35]{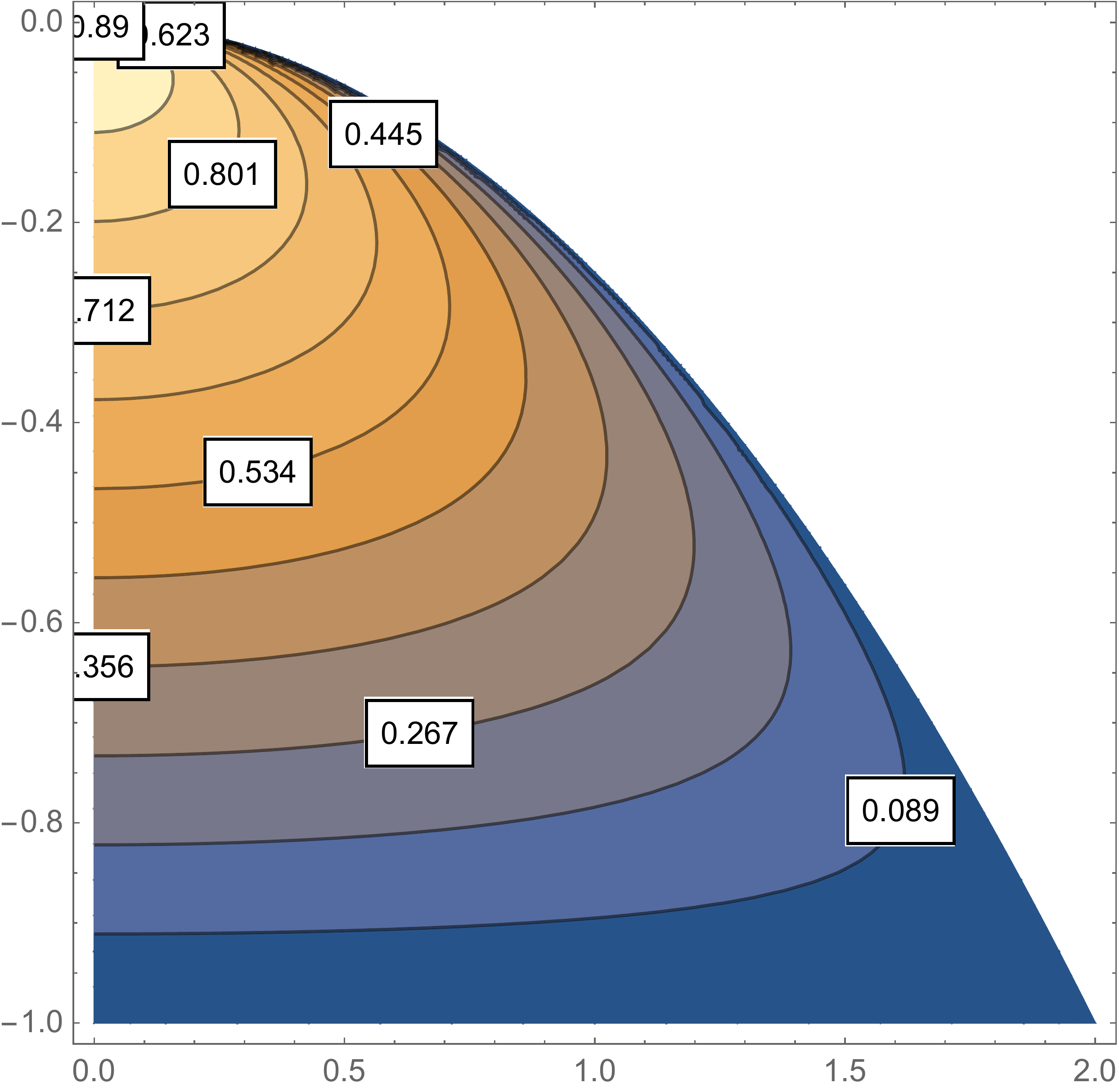}
\caption{Contour plot of the square of the gap in the region $0\le \lambda_1\le 2$ and $0\ge \lambda_2 \ge -1$}
\label{fig:Contour} 
\end{center}
\end{figure}

\begin{enumerate}
\item For the Ising model in a transverse field without  three spin interaction,
the gaps collapse at the Brillouin zone boundaries, $k=\pm \pi$ at the self-dual point $\lambda_{1}=1$ and $\lambda_{2}=0$.
\item As we move along the critical line $\lambda_{2}=1-\lambda_{1}$, there are no additional critical points until we reach $\lambda_{2}=1$ , then  the   gaps collapse at $k=0$ when $\lambda_{1}=0$ and $\lambda_{2}=1$, In the language of conformal field theory, this crossover is described by Zamolodchikov’s c-function’, which takes one from a theory of conformal charge $c=1$ to that of charge $c=1/2$. \cite{Neto:1993} Then on $\lambda_{2}=1+\lambda_{1}$ constitutes a critical line with   criticality at $k=0$.
\item When the three spin interaction is added, the gaps
also collapse at  $k=\cos^{-1}(\lambda_{1}/2)$ for  $\lambda_{2}=-1$ and $0<\lambda_{1}<2$. This constitutes an unusual incommensurate critical line.
At exactly $\lambda_2=-1$ and $\lambda_1=2$, the multicritical point is no longer Lorentz invariant and the dynamical exponent is $2$. This has important consequence in the spin-spin correlation function didcussed below because conformal invariance no longer holds.
\item In the  dual representation discussed below, the region enclosed by $\lambda_{1}^{2}=-4\lambda_{2}$ is an oscillatory ferromagnetically ordered phase separating
from an ordered phase for $\lambda_{2}<0$, as determined by the spatial decay of the instantaneous spin-spin correlation function. This does not reflect a critical line, however.\cite{Barouch:1971}
\item It is easy to compute specific heat, $C$, which is linearly proportional to $T$, as $T\to 0$, on all critical lines except at the multicritical point $(d)$ in Fig.~\ref{fig:fig1}; $C$ vanishes as $T^{1/2}$. The theory at this point is not conformally invariant. Disorder has a strong effect at this point as well, as we shall see later.
\end{enumerate}

\begin{figure}[htbp]
\begin{center}
\includegraphics[scale=0.35]{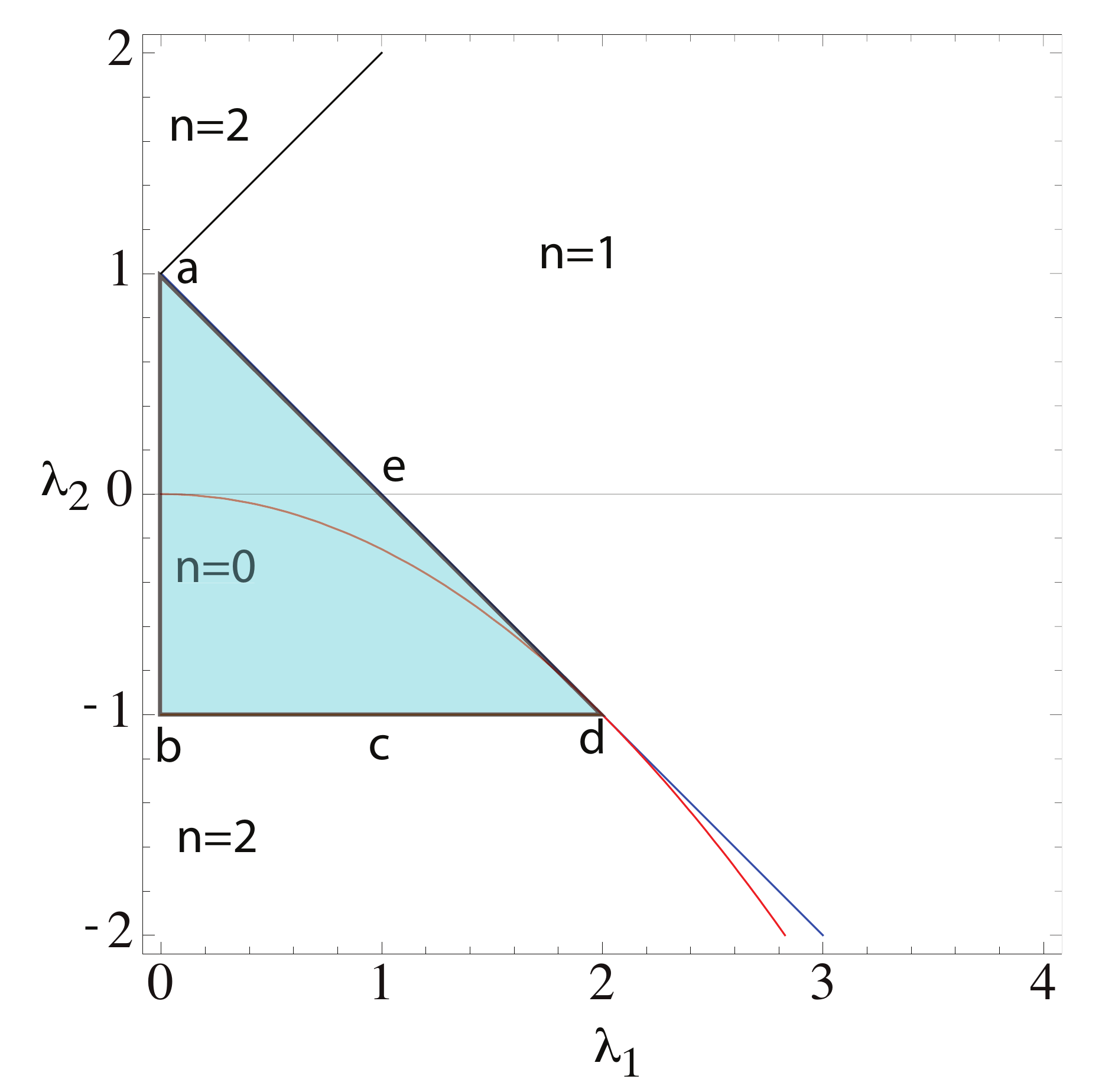}
\caption{The phase diagram, $n=0, 1, 2$ correspond to  regions with $n$ Majorana zero modes at each end of an open chain. The quantum critical lines are determined by the vanishing of the energy gaps. The points (a) and (d) are multicritical points. The thin green line is a disorder line of no significance to quantum criticality.~\cite{Niu:2012}}
\label{fig:fig1}
\end{center}
\end{figure}

\begin{figure}[htbp]
\begin{center}
\includegraphics[scale=0.3]{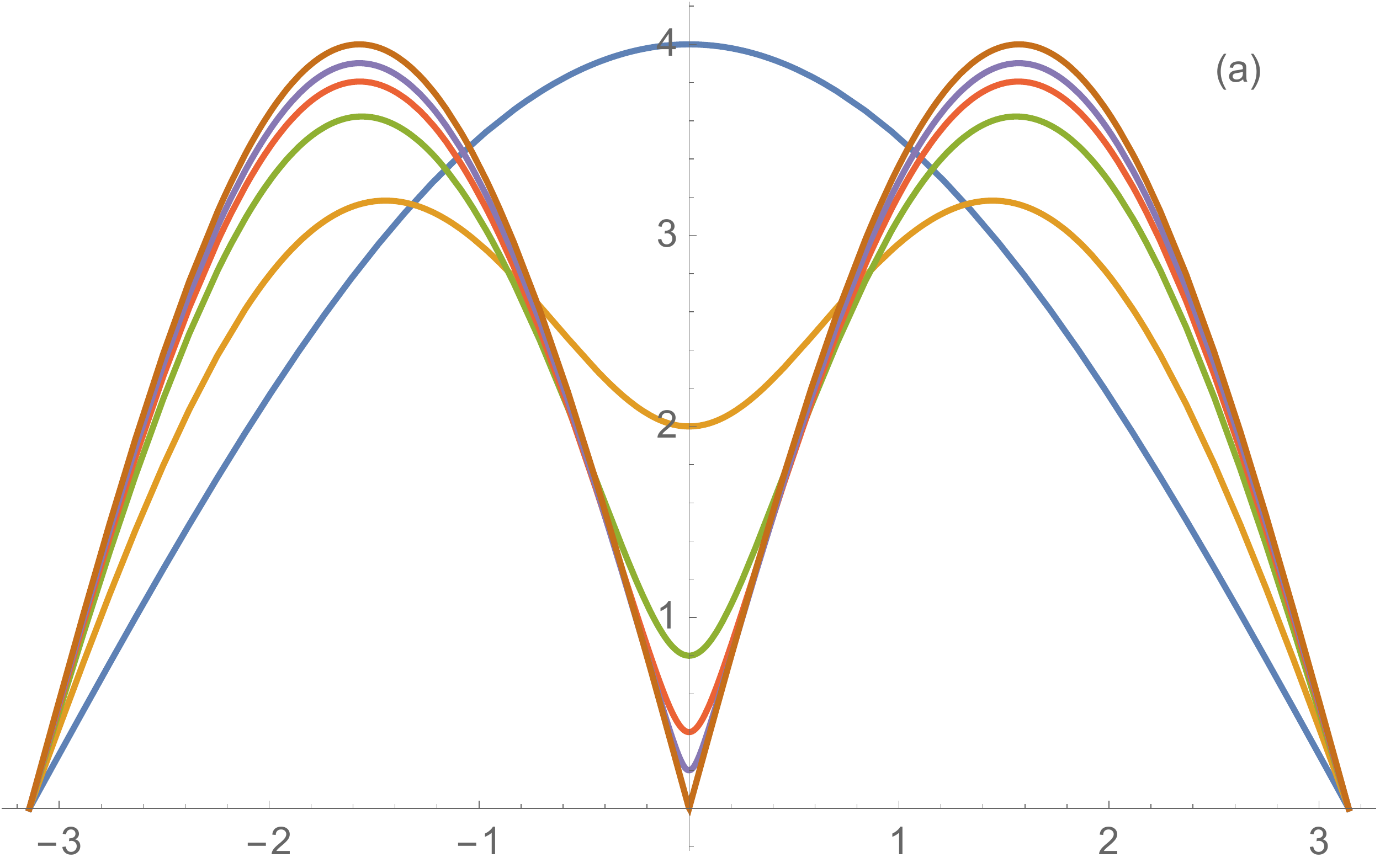}
\includegraphics[scale=0.3]{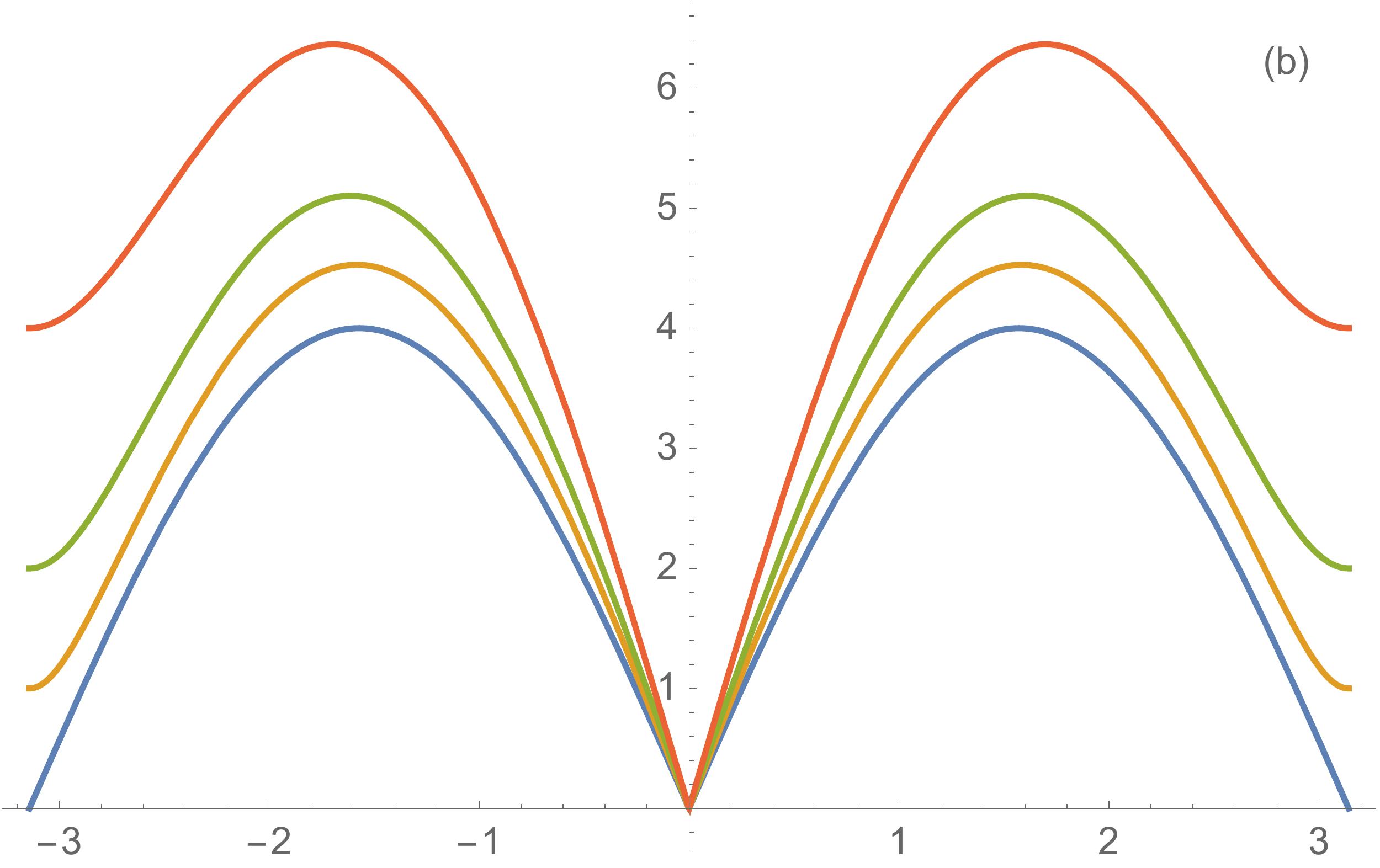}
\includegraphics[scale=0.3]{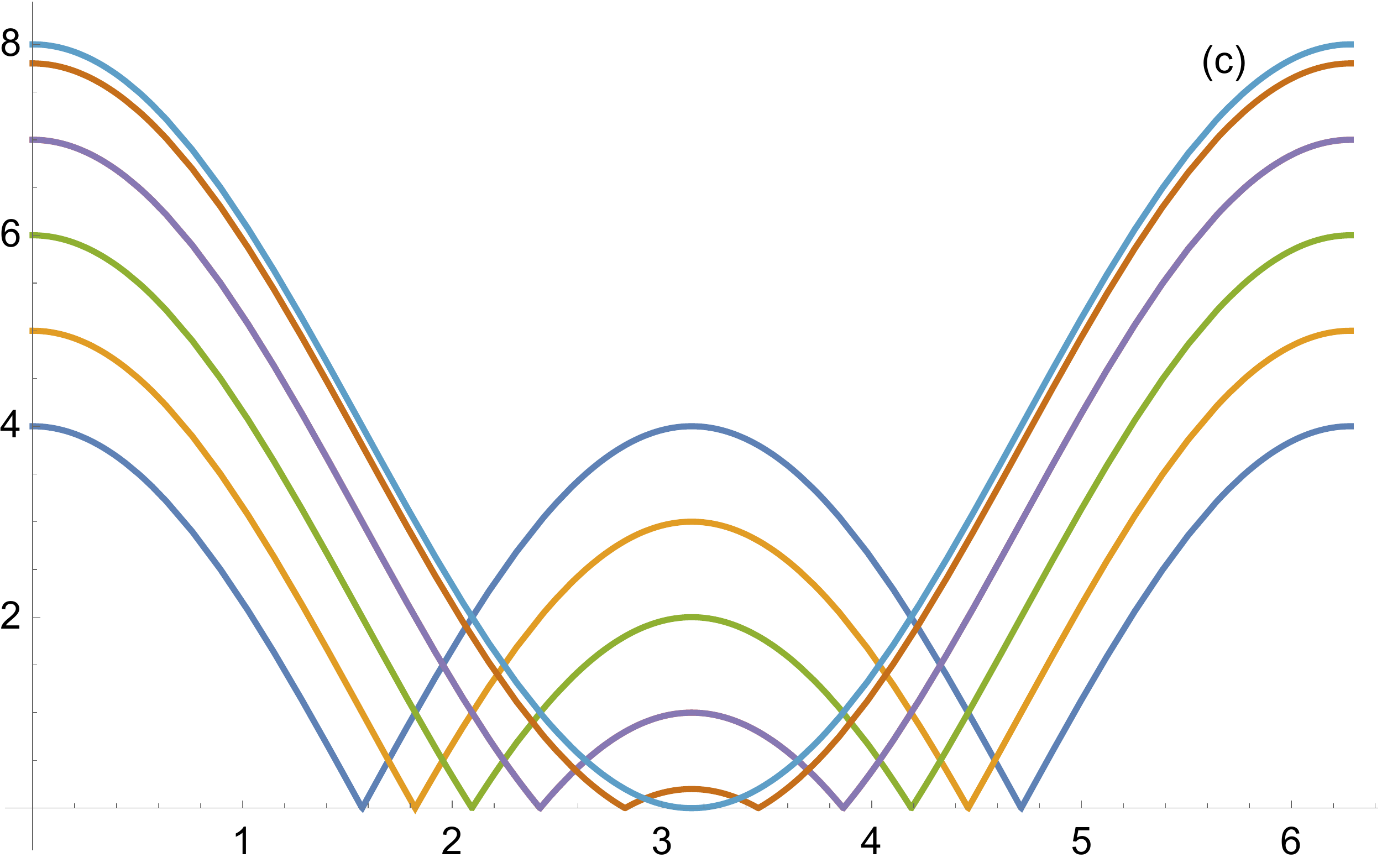}
\caption{2$E_k$ versus momentum $k$. The critcality along (a) $\lambda_{2}=1-\lambda_{1}$: top to bottom $\lambda_1\to 0$, (b)  $\lambda_{2}=1+\lambda_{1}$ where $\lambda_1$ varies between $0$ and $1$,
(c) $\lambda_2=-1, 0\le \lambda_1 \le 2$, plotted in the interval between $0$ and $2\pi$.  }
\label{fig:fig2}
\end{center}
\end{figure}

 But at the free fermion point $b$,  there are no zero energy excitations except at $k=\pm \pi/2$. When we move to the point $c$ the spectrum evolves increasing  the weight at $k=0$ and the locations of the nodes are incommensurate with the lattice. The incommensuration shifts as a function of $\lambda_{1}$. The point $d$ 
 is a multicritical point, with the dynamical critical  point $z=2$ and the spectra vanishes quadratically at $\pm \pi$. The spectra are no longer relativistic at this point as a result of the confluence of two Dirac points,  corresponding to a dynamical exponent $z=2$ and the correlation length exponent $\nu=1/2$.  The multicritacl point at $a$ still corresponds to dynamical exponent $z=1$, but its location is now at $k=0$. The linearly vanishing spetrum continues as $\lambda_1=1+\lambda_2$ located at $k=0$. 

\subsection{The dual representation}
In this section we exhibit the dual representation, which shows that the model is also dual to an anistropic $XY$-model with a magnetic field in the $z$-direction. We show the transcription in case the  $XY$-version  is better realized from tht possible experimental perspective.

Let us define the dual operators:
\begin{eqnarray}
\mu_x(n) &=& \sigma_z(n+1)\sigma_z(n),\\
\mu_z(n)&=& \prod_{m\le n} \sigma_x(m),
\end{eqnarray}
which implies that 
\begin{eqnarray}
[\mu_z(n),\mu_x(n)]&=& -2i\mu_y(n),\\
\mu_y(n)&=&-i \bigg(\prod_{m \le n}\sigma_x(m)\bigg) \sigma_z(n+1)\sigma_z(n).
\end{eqnarray}
The Hamiltonian under  duality transforms to
\begin{equation}
\begin{split}
H_D&=-\frac{2}{1+r}\sum_n\bigg[\frac{1+r}{2}\mu_x(n)\mu_x(n+1)\\&+\frac{1-r}{2}
\mu_y(n)\mu_y(n+1)+h\mu_z(n))\bigg],\label{Hxy}
\end{split}
\end{equation}
where we have carried out the rotations : $\mu_x(n)\to \mu_z(n)$, $\mu_z(n)\to \mu_x(n)$, $\mu_y(n)\to -\mu_y(n)$.  The   parameters are related  by
\begin{equation}
\lambda_1 = \frac{2h}{1+r}, \;
\lambda_2 = \frac{r-1}{1+r}.
\end{equation}
The critical line in the $XY$-model, separating the quantum disordered phase from the ferromagnetic phase is at $h=1$, which corresponds to
$\lambda_1+\lambda_2=1$,
separating the  ordered phase  from the disordered phase. Since the ordered and the disordered phases are exchanged under duality,    the disordered phase of
the three-spin model is $\lambda_1+\lambda_2<1$.  

\section{Majorana representatrion}
The phase transitions in this model in the superconducting version (in the fermion language)  is best described in terms of Majorana zero modes at the end of the one-dimensional chain. This view was discussed in detail in a previous paper.~\cite{Niu:2012} While in the spin language the phases are described in the traditional language of an order parameter, ferromagnetic or quantum disordered states, in the fermion language, the system is always a superconductor, except that the states are  either  BCS or BEC phases. That there is a phase transition between the two is a reflection of the topological nature of the states. Referring to  Fig.~\ref{fig:fig1}, the winding numbers $n=0,1, 2$ are the number of zero energy Majorana modes at  each end of the chain. These windings were explained in terms of the Anderson pseudospin Hamiltonian,~\cite{Anderson:1958} when the time reversal symmetry is respected. In the absence of time reversal symmetry, the two Majorana zero modes can hybridize and split.  The change in the winding numer cannot take place without a phase transition.  More recently, the topological nature was brought out by the notion of curvature renormalization group (CRG)\cite{Sarkar:2021,Rao:2020} that may be useful in higher dimensional systems. Here there is an exact solution of the model.

\section{Dynamical structure factor}
We examine both pure and disordered versions at  zero temperature.  For this purpose we compute spin-spin correlation function, which is possible to determine by Neutron scattering experiments.. It is difficult to find a suitable experimental technique in the superconducting picture, except perhaps by scanning tunneling microscopy---a topic that we would like to return in the future. The dynamical structure factor $S(k,\omega)$ can be measured in neutron scattering experiment. It is fortunate that we can provide a numerically exact method in this respect. Alternately one can also explore NMR relaxation rate. 

For a pure system, $S(k,\omega)$ it contains no more information beyond the results for the excitation spectrum, but it encourages the experimentalists to perform the neutron measurements by measuring the spin-spin correlation function. Secondly, the method of calculation of the correlation function for the pure system can be tested before we consider the more complex disordered systems. Another important point is that it is rare that one can compute the real-time correlation function. In the present problem, this is possible because we have the full exact spectrum regardless of whether or not we have a pure or disordered case.

We treat two different models of disorder: the binary and the uniform distributions. The binary distribution consists of a large field $h_{L}$ and a small field $h_{S}$, with the probability distributions such that $P_{L} + P_{S}=1$ where
\begin{equation}
h_{i}=\begin{cases}
               h_{L} &\text{with probability $ P_L $}\\
               h_{S} &\text{with probability $ P_S$}
            \end{cases} 
\end{equation}            
in this case we keep $\lambda_1$ and $\lambda_2$ spatially independent. 
The probability distribution $P(h)$ for the binary distribution  is
\begin{equation}
P(h)=P_{L} \ \delta(h-h_{L})+P_{S} \ \delta(h-h_{S}) 
\end{equation}

The uniform distribution for $h$ is given by its average and the width
\begin{equation}
P(h_i)=\begin{cases}
             1/h_W, & h_i \in [h_{ave} - \frac{1}{2}h_W,h_{ave} + \frac{1}{2}h_W] \\
             0, &\text{otherwise. }
             \end{cases}
\end{equation}

\subsection{Correlation function in terms of a Pfaffian}
We discuss the calculations   in some detail for the reader to be able to repeat the calculation with ease, if necessary. Quite generally, the spin-spin correlation function $C_{ij}(t)$ is defined as 
\begin{equation}
C(r,t)=\left<\sigma_{i}^{z}(t)\sigma_{j}^{z}(0)\right>
\end{equation}
where $i$, $j$ are lattice sites and $r$ is the separation between them. The dynamical structure factor $S(k,\omega)$ is the time and space Fourier transformation of $C_{ij}(t)\equiv C(j-i, t)\equiv C(r,t)$. In a finite system of length $L$, we choose $i,j$ in the middle of the chain to reduce the boundary effects. Thus,
\begin{equation}
S(k,\omega) = \int dt \int dr \ e^{i\omega t}e^{-ikr}C(r,t).
\end{equation}
The integral over $r$ represents of course a discrete sum on a lattice. Furthermore, for the disordered case we compute 
\begin{equation}
{\overline C(r,t)}=\overline{<\sigma_{i}^{z}(t)\sigma_{j}^{z}(0)>}
\end{equation}
where the overline stands for an average over the disorder ensemble. The dynamical structure factor $S(k,\omega)$ is 
\begin{equation}
S(k,\omega) = \int dt \int dr \ e^{i\omega t}e^{-ikr}C(r,t)
\end{equation}
Using the Jordan-Wigner transformation, Eq.~\ref{Eq:JW-2},
we get\cite{Lieb:1961}
\begin{widetext}
\begin{equation}
C(r,t)=\left<(\prod_{m=1}^{i-1}(c_{m}^{\dagger}+c_{m})(t)(c_{m}^{\dagger}-c_{m})(t))(c_{i}^{\dagger}+c_{i})(t)(\prod_{l=1}^{j-1}(c_{l}^{\dagger}+c_{l})(c_{l}^{\dagger}-c_{l}))(c_{j}^{\dagger}+c_{j})\right>
\label{Eq:Correlation}
\end{equation}
\end{widetext}

Because of the free fermion nature of the Jordan-Wigner transformed Hamiltonian, we can apply Wick's theorem to $C(r,t)$. After collecting all terms in the Wick expansion, this gives us a Pfaffian.
It means that
\begin{equation}
\left<\sigma_{i}^{z}(t)\sigma_{j}^{z}(0)\right> =   Pf(S) 
\end{equation}
Here $S$ is a $2(i+j-1)$ dimensional skew-symmetric matrix. If we identify $A_{m} = c_{m}^{\dagger}+c_{m}$ and $B_{n} = c_{n}^{\dagger}-c_{n}$ The dmatrix $S$ is 
\begin{widetext}
\begin{equation}
S =
\begin{pmatrix}
0 & <A_{1}(t)B_{1}(t)> & <A_{1}(t)A_{2}(t)> & <A_{1}(t)B_{2}(t)> & ... & <A_{1}(t)A_{j}> \\
-<A_{1}(t)B_{1}(t)> & 0 & <B_{1}(t)A_{2}(t)> & <B_{1}(t)B_{2}(t)> & ... & <B_{1}(t)A_{j}>\\
-<A_{1}(t)A_{2}(t)> & -<B_{1}(t)A_{2}(t)> & 0 & <A_{2}(t)B_{2}(t)> & ... &
<A_{2}(t)A_{j}>\\
\vdots & \vdots & \vdots & \vdots & \ddots & \vdots \\
-<A_{1}(t)A_{j}> & -<B_{1}(t)A_{j}>  & -<A_{2}(t)A_{j}> & -<B_{2}(t)A_{j}> & ... & 0
\end{pmatrix}
\end{equation}
\end{widetext}
All we need to compute is the two point correlation function\cite{Fetter:1971} such as
\begin{equation}
\left<(c_{m}^{\dagger}(t)\pm c_{m}(t))(c_{l}^{\dagger}\pm c_{l})\right>.
\end{equation}

Here are the details of the remaining part of the calculation.. The matrix form of the Hamilitonian  in the fermion basis is 
\begin{equation}
H = \begin{matrix}\begin{pmatrix}c^{\dagger} & c\end{pmatrix}\end{matrix}
  \begin{pmatrix} A & B \\ -B & -A \end{pmatrix} 
  \begin{pmatrix} c \\c^{\dagger}  \end{pmatrix}
\end{equation}
in which $c = (c_{1} , c_{2} ..... c_{L})$, $c^{\dagger}= (c_{1}^{\dagger} , c_{2}^{\dagger} ..... c_{L}^{\dagger})$, $A$ and $B$ are both $L\times L$ matrices, whose elements  are 
\begin{eqnarray}
A_{ij} &=& \lambda_{1}(\delta_{j,i+1}+\delta_{j,i-1})+\lambda_{2}(\delta_{j,i+2}+\delta_{j,i-2})-2h_{i}\delta_{ij}\\
B_{ij}&=& -\lambda_{1}(\delta_{j,i+1}-\delta_{j,i-1})-\lambda_{2}(\delta_{j,i+2}-\delta_{j,i-2})
\end{eqnarray}
Now,  we diagonalize $H$ as 
\begin{equation}
H = \begin{matrix}\begin{pmatrix}c^{\dagger} & c\end{pmatrix}\end{matrix}
   VDV^{-1}
  \begin{pmatrix} c \\c^{\dagger}  \end{pmatrix}
  = \begin{matrix}\begin{pmatrix}\eta^{\dagger} & \eta\end{pmatrix}\end{matrix}
  D
  \begin{pmatrix} \eta \\\eta^{\dagger} \end{pmatrix}
\end{equation}
Here $D$ and $V$ are $2L\times 2L$ matrices that take the following forms:
\begin{equation}
 D = \begin{bmatrix}
    E_{1} &  \\
     & E_{2} &  \\
     &  & E_{3}  \\
     &  & &  ...\\
     &  & & &  ...\\
     &  & & &   & E_{2L}
\end{bmatrix} , 
\end{equation}
    \begin{equation}
V = \begin{pmatrix} X_{L\times L} & Y_{L\times L} \\ Y_{L\times L} & X_{L\times L} \end{pmatrix} 
\end{equation}
$(X)_{L\times L}$ means a $L\times L$ matrix. Now, our Hamilionian in the basis of $\eta$ and $\eta^{\dagger}$ above should becomes
\begin{equation}
H=\sum_{\mu=1}^{L} 2E_{\mu}\left(\eta_{\mu}^{\dagger}\eta_{\mu}-1\right).
\end{equation}
We represent everything in real-space since momentum representation in Eq.(9) doesn't work in the disorder case; $\mu$ is no longer  the momentum quantum number. The first term is the  diagonalized Hamilitonian and the second term is the ground state energy.
Next, in order to evaluate Eq.(28), we need to rewrite $c^{\dagger}$ and $c$ in terms of $\eta^{\dagger}$ and $\eta$. Since $\begin{pmatrix}
c  \\
c^{\dagger}
\end{pmatrix} = 
V\begin{pmatrix}
\eta \\
\eta^{\dagger}
\end{pmatrix}$, we have
\begin{equation}
 \begin{pmatrix}
c^{\dagger}+c  \\
c^{\dagger}-c 
\end{pmatrix} = 
\begin{pmatrix}
X+Y & 0  \\
0 & X-Y
\end{pmatrix} \begin{pmatrix}
\eta^{\dagger}+\eta  \\
\eta^{\dagger}-\eta 
\end{pmatrix}.
\end{equation}
Evaluation of Eq.(27) requires all possible $m$ and $l$ in Eq.(28). A nice way to do this is to rewrite Eq.(28) as a $2L\times 2L$ matrix called $C_{t}$ whose elements contain all possible two points correlators. 
\begin{equation}
C_{t} =
\left< \begin{pmatrix}
c^{\dagger}(t)+c(t) \\
c(t)-c^{\dagger} (t)
\end{pmatrix} \begin{pmatrix}
c^{\dagger}+c & c-c^{\dagger} 
\end{pmatrix}  \right> 
\end{equation}
Here $c+c^{\dagger}
\equiv (c_{1}+c_{1}^{\dagger} , c_{2}+c_{2}^{\dagger}, ..... ,c_{L}+c_{L}^{\dagger})$. Then we use Eq.(36) to get 
\begin{widetext}
\begin{equation}
\left< \begin{pmatrix}
c^{\dagger}(t)+c(t) \\
c(t)-c^{\dagger} (t)
\end{pmatrix} \begin{pmatrix}
c^{\dagger}+c & c-c^{\dagger} 
\end{pmatrix} \right > = \begin{pmatrix}
X+Y & 0  \\
0 & X-Y
\end{pmatrix} 
\left< \begin{pmatrix}
\eta^{\dagger}(t)+\eta(t) \\
\eta(t)-\eta^{\dagger} (t)
\end{pmatrix} \begin{pmatrix}
\eta^{\dagger}+\eta & \eta-\eta^{\dagger} 
\end{pmatrix}\right > 
\end{equation}
$$
\begin{pmatrix}
X^{\dagger}+Y^{\dagger} & 0  \\
0 & X^{\dagger}-Y^{\dagger}
\end{pmatrix} 
$$
\end{widetext}
Now We just need to evaluate $\left< \begin{pmatrix}
\eta^{\dagger}(t)+\eta(t) \\
\eta(t)-\eta^{\dagger} (t)
\end{pmatrix} \begin{pmatrix}
\eta^{\dagger}+\eta & \eta-\eta^{\dagger} 
\end{pmatrix}  \right>$. 

We calculate that by first computing $\left<\begin{pmatrix}
\eta(t) \\
\eta^{\dagger}(t)
\end{pmatrix} \begin{pmatrix}
\eta^{\dagger} & \eta
\end{pmatrix}\right>$. Here $\eta(t) \equiv (\eta_{1}(t), \eta_{2}(t), ..... ,\eta_{L}(t))$. 
Fortunately, the only non-zero contraction is between $\eta_{i}(t)$ and $\eta_{j}^{\dagger}$. It gives 
$\left<\eta_{i}(t)\eta_{j}^{\dagger}\right> = \delta_{ij}e^{i2E_{i}t}$, 
Now our matrix $C_{t}$ becomes
\begin{widetext}
\begin{equation}
C_{t} =
\begin{pmatrix}
X+Y & 0  \\
0 & X-Y
\end{pmatrix} \begin{pmatrix}
(\delta_{ij}e^{i2E_{i}t})_{L\times L} & -(\delta_{ij}e^{i2E_{i}t})_{L\times L}  \\
-(\delta_{ij}e^{i2E_{i}t})_{L\times L} & (\delta_{ij}e^{i2E_{i}t})_{L\times L}
\end{pmatrix} \begin{pmatrix}
X^{\dagger}+Y^{\dagger} & 0  \\
0 & X^{\dagger}-Y^{\dagger}
\end{pmatrix} 
\end{equation}
\end{widetext}
Then, if one finds $X$ and $Y$ correctly, we are able to compute $C_{t}$ without any difficulty. However, getting those eigenvectors $X$ and $Y$ from exact diagonalization of $H$ is not numerically stable (suffers large errors) if their eigenvalues are very close to zero. Thus, instead of directly diagonalizing our $2L\times2L$ hamilitonian, we choose to use singular value decomposition (SVD)\cite{Press:2007} to diagonalize $L\times L$ matrix.  To this end, we change our basis from $c$ and $c^{\dagger}$ into $c^{\dagger}+c$ and $c^{\dagger}-c$; our Hamilitonian transforms to 
\begin{equation}
\widetilde{H} =\begin{pmatrix}c^{\dagger}+c & c-c^{\dagger}\end{pmatrix}
   \begin{pmatrix}
0 & M^{T} \\
M & 0
\end{pmatrix} 
\begin{pmatrix} c^{\dagger}+c \\c^{\dagger}-c  \end{pmatrix}
\end{equation}
in which $M = -\frac{(A+B)}{2}$ and $M^{T} = -\frac{(A-B)}{2}$. 

The exact form of $M$ is 
\begin{equation}
 M = \begin{bmatrix}
    h_{1} & -\lambda_{1} & -\lambda_{2} &  &  \\
     & h_{2} & -\lambda_{1} & -\lambda_{2}  & & \\
     &  & h_{3} & -\lambda_{1} & \ddots & \\
     &  & & \ddots & \ddots & -\lambda_{2} \\
     &  & & & \ddots & -\lambda_{1} \\
     &  & & &   & h_{L} 
\end{bmatrix}
\end{equation}
Next we apply SVD to matrix $M$. This gives us
\begin{equation}
M = \psi \Lambda \phi^{T}   
\end{equation}
where $\psi$ and $\phi$ are $L \times L$ orthogonal matrices and $\Lambda$ is diagonal with non-negative entries. The SVD also implies 
\begin{equation}
M \phi = \psi \Lambda    
\end{equation}
\begin{equation}
M^{T} \psi = \phi \Lambda    
\end{equation}
Those are equivalent to the equation 
\begin{equation}
\widetilde{H}\begin{pmatrix}
\phi & \phi  \\ 
\psi & -\psi
\end{pmatrix} = \begin{pmatrix}
\phi & \phi  \\ 
\psi & -\psi
\end{pmatrix} \begin{pmatrix}
\Lambda & 0  \\ 
0 & -\Lambda
\end{pmatrix}
\end{equation}
Then one can show the equivalence between $X+Y$, $X-Y$
and $\phi$, $\psi$ in SVD.  We have 
\begin{equation}
\begin{pmatrix}
\phi & 0  \\
0 & \psi
\end{pmatrix} =\begin{pmatrix}X+Y & 0  \\
0 & X-Y
\end{pmatrix}
\end{equation}
Finally, our two point fermion-fermion correlation function becomes 
\vspace {0.5cm}
\begin{widetext}
\begin{equation}C_{t} 
 = \left< \begin{pmatrix}
c^{\dagger}(t)+c(t) \\
c(t)-c^{\dagger} (t)
\end{pmatrix} \begin{pmatrix}
c^{\dagger}+c & c-c^{\dagger} 
\end{pmatrix} \right> = \begin{pmatrix}
\phi & 0  \\
0 & \psi
\end{pmatrix} \begin{pmatrix}
(\delta_{ij}e^{i2E_{i}t})_{L\times L} & -(\delta_{ij}e^{i2E_{i}t})_{L\times L}  \\
-(\delta_{ij}e^{i2E_{i}t})_{L\times L} & (\delta_{ij}e^{i2E_{i}t})_{L\times L}
\end{pmatrix} \begin{pmatrix}
\phi^{\dagger} & 0  \\
0 & \psi^{\dagger} 
\end{pmatrix} 
\end{equation}
\end{widetext}
Now, we can get all elements in matrix $S$ by mapping to all elements in $C_{t}$. But we still need to deal with one last issue. The computation of a Pfaffian consumes a lot of time by standard methods for a large-sized system. An efficient method for calculating such Pfaffians was invented  in a previous paper.\cite{Jia:2006}  Let $X$ be a $2N \times2 N$ skew-symmetric matrix which has  the following form
\begin{equation}
X=\left[ {\begin{array}{cc}
   A & B \\
   -B^{T} & C \\
  \end{array} } \right]
\end{equation} 
where $A$ is a $2\times 2$ matrix, and $B$ and $C$ are matrices of appropriate dimensions.Then we have the identity
\begin{equation}
\left[ {\begin{array}{cc}
   I_{2} & 0 \\
   B^{T}A^{-1} & I_{2N-2} \\
  \end{array} } \right]X\left[ {\begin{array}{cc}
   I_{2} & -A^{-1}B \\
   0 & I_{2N-2} \\
  \end{array} } \right]
  =\left[ {\begin{array}{cc}
   A & 0 \\
   0 & C+B^{T}A^{-1}B\\ 
  \end{array} } \right]
\end{equation}   
where $I_{n}$ is a $n\times n$ identity matrix, and
\begin{equation}
\mathrm{det}(X)=\mathrm{det}(A)\mathrm{det}(C+B^{T}A^{-1}B)
\end{equation}
This  gives us an iteration method. We will get a $2\times2$ matrix $A$ in each iteration step;  we treat $C+B^{T}A^{-1}B$ to be our next $X$ and keep doing this. Our $\det(X)$ eventually becomes a product chain of $2\times2$ matrices.

\subsection{Pure System}

\begin{figure}[htbp]
\includegraphics[scale=0.4]{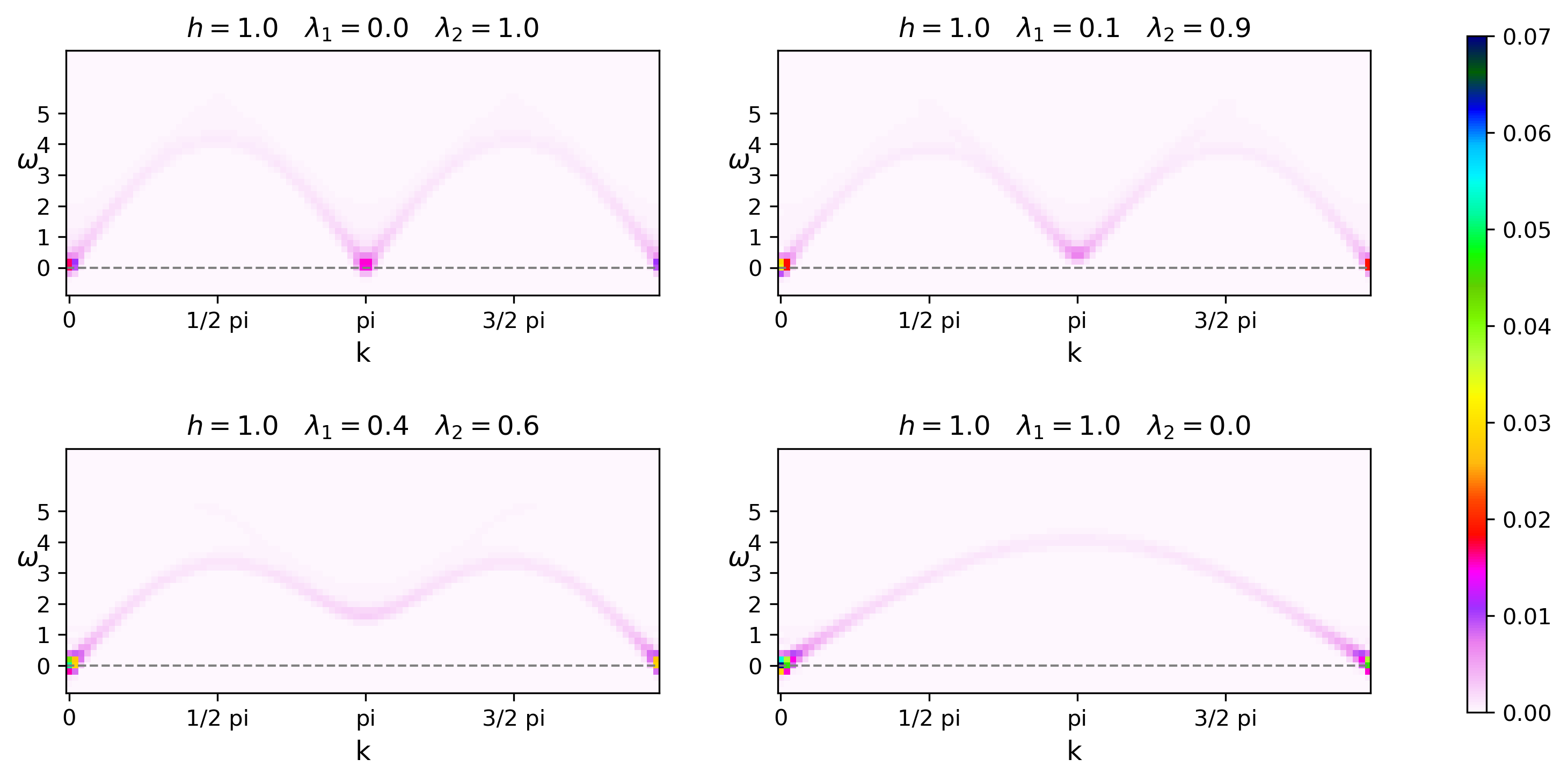}
\centering
\caption{Normalized dynamical structure factor $S(k,\omega)$  calculated at four  points..}
\label{fig:pure-1}
\end{figure}

\begin{figure}[htbp]
\includegraphics[scale=0.35]{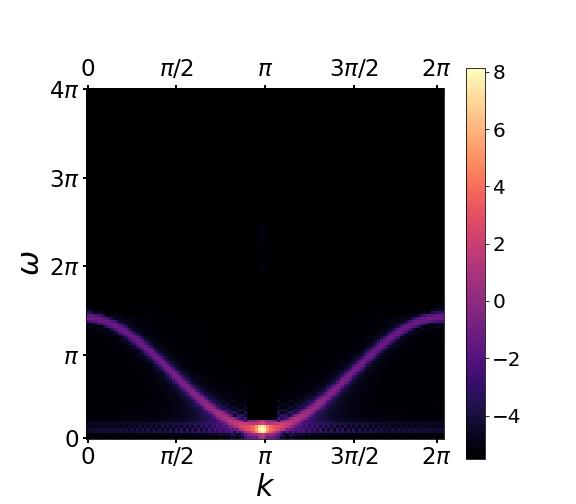}
\includegraphics[scale=0.35]{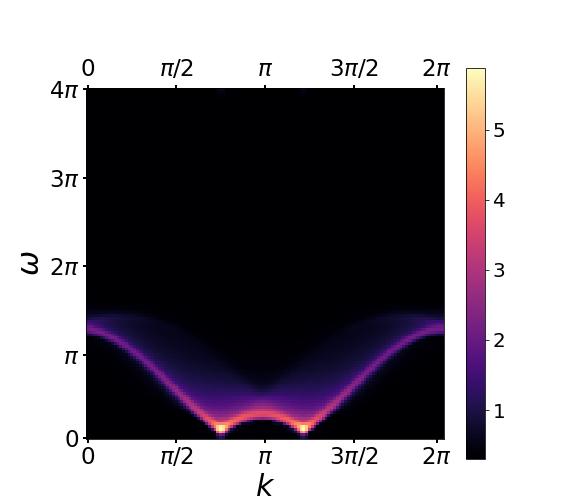}
\centering
\caption{Top: Dynamical structure factor $S(k,\omega)$  calculated  at the multicriitical point $\lambda_2=-1$ and $\lambda_1=2$; Bottom:.$\lambda_2=-1$ and $\lambda_1=1.5$.}
\label{fig:pure2}
\end{figure}

The calculations in this section were performed on a chain that has 128 lattice sites with free boundary conditions. We always choose 64 sites in the middlke to compute the correlation function.
In the pure model, we focus on how dynamical structure factor $S(k,\omega)$ evolves along the line $\lambda_{2} = 1 \pm\lambda_{1}$ as represented in Fig.~\ref{fig:pure-1}. In addition, we also plot the result as we tune $\lambda_1$ to the multicritical point $\lambda_1=2$ and $\lambda_2=-1$; see Fig.\ref{fig:pure2}. Possible neutron scattering will  measure spin-spin correlation function. It is satisfying to see that the spectra still correctly represent the fermionic excitations. It is of course not possible to directly couple to the Jordan-Wigner fermions. In addition it is a good test that our calculations are reliable and we can safely continue to the disorder problem

\subsection{Disordered System}
A cursory look at  Eq.~\ref{eq:Ham1}  would seem that the disorder in this one dimensional model will result in Anderson localization of the fermionic states. This not true because the Hamltonian contains pair creation and destruction operators; it corresponds to a model of a superconductor, in which charge is not conserved. 

\begin{figure}[htbp]
\includegraphics[scale=0.35]{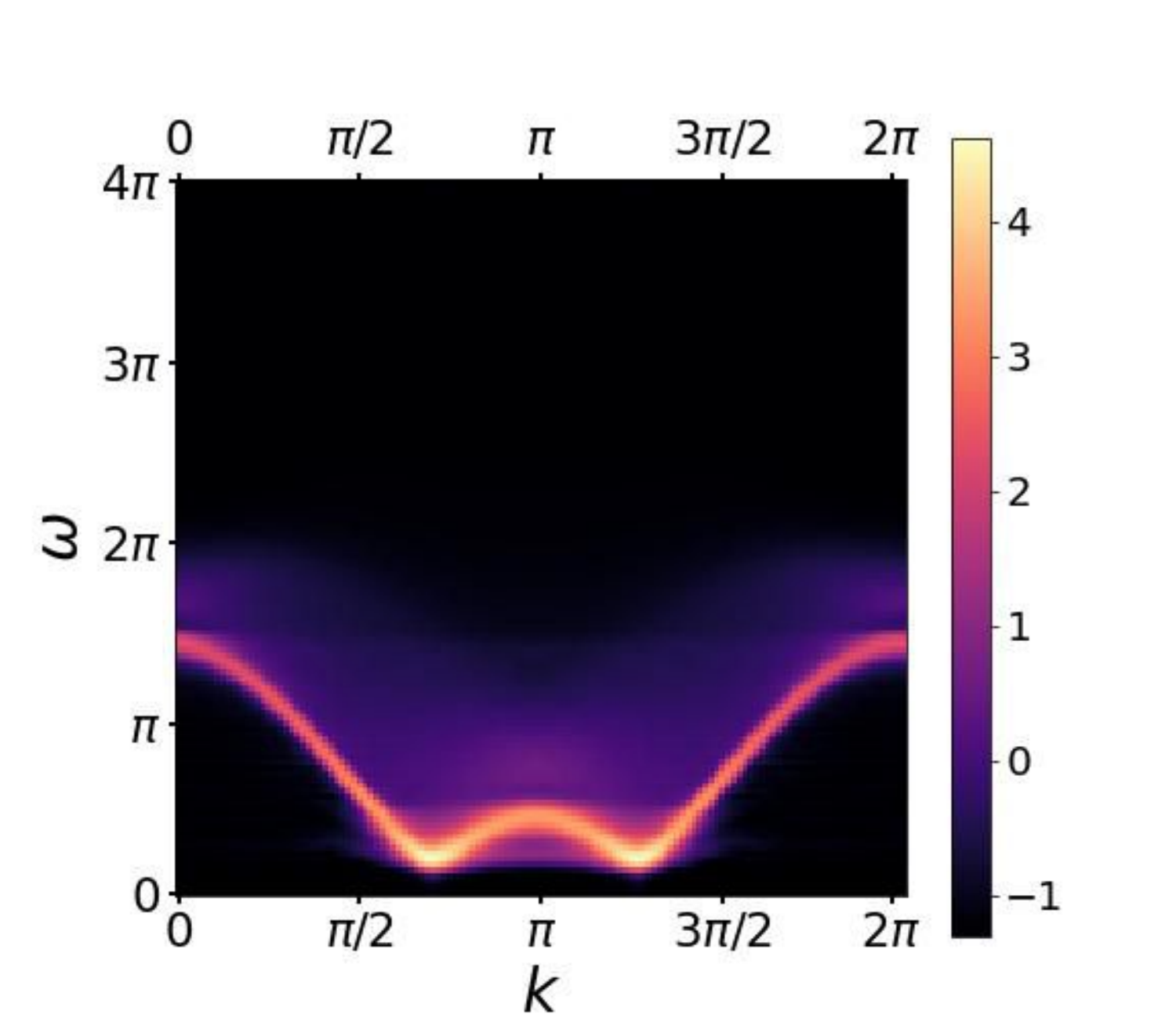}
\includegraphics[scale=0.35]{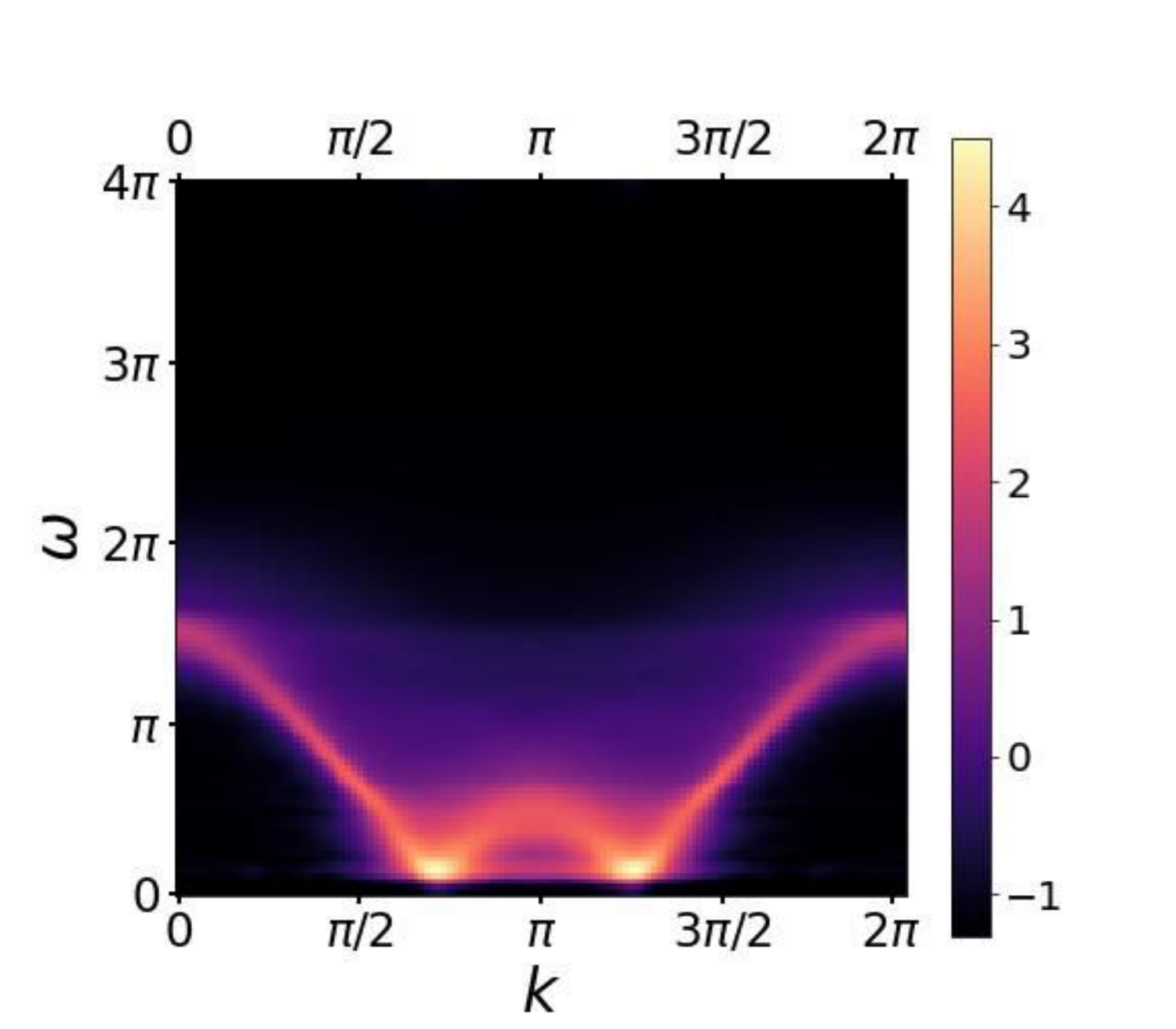}
\caption{Dynamical structure factor $S(k,\omega)$ calculated for a uniform disorder. Top:  $\lambda_2=-1$ and $\lambda_1=1.5, h_w=1$; Bottom:  $h_w=2$.}
\label{fig:uniform}
\end{figure}

\begin{figure}[htbp]
\includegraphics[scale=0.35]{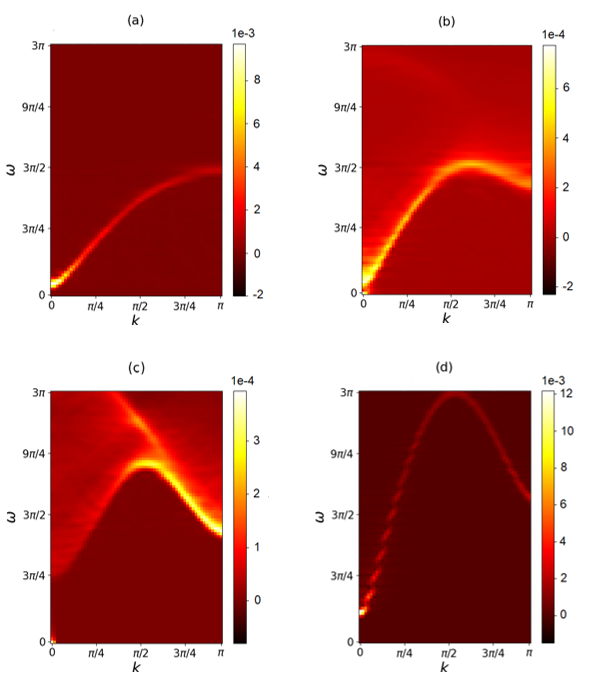}
\caption{Normalized dynamical structure factor $S(k,\omega)$ with
increasing value of $\lambda_2$ for uniform disorder. The parameters are $\lambda_1=1$, $h_{av}=1.4$ and $h_{w} = 0.5$:  $ \lambda_2= (a)\; 0.1, 
(b)\; 0.5, (c)\;1.0, 
(d)\;3.0$. }
\label{fig:disorder2}
\end{figure}

\begin{figure}[htbp]
\includegraphics[scale=0.35]{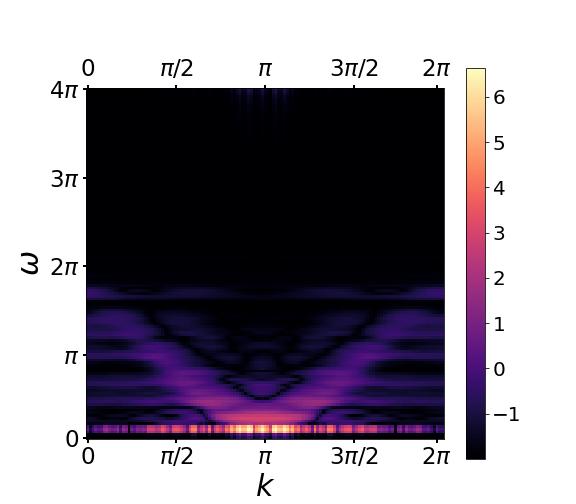}
\caption{The parametersare $\lambda_1=2,\lambda_2=-1,h_{L}=3$, $h_{S}=1$, $P_{L}=0.1$ at the multicritical point. The quadratic dispersion is still visible.}
\label{fig:disorder1}
\end{figure}

\begin{figure}[htbp]
\includegraphics[scale=0.35]{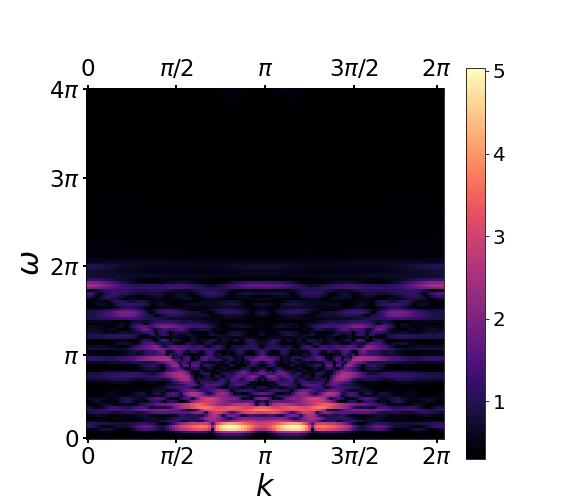}
\caption{The parameters are $\lambda_1=2,\lambda_2=-1,h_{L}=3$, $h_{S}=1$, $P_{L}=0.5$. The dispersion is broken up for large disorder.}
\label{fig:disorder3}
\end{figure}

Computation of $S(k,\omega)$ involves averaging of the correlation function.
We found that there appears to be no difference between  configuration averages of 1000 and 3000 samples. A representative set of $S(k,\omega)$ are shown in Figs.~\ref{fig:uniform},\ref{fig:disorder2},\ref{fig:disorder1},\ref{fig:disorder3}.
This example corresponds to little disorder and therefore very little broadening and the dispersions remains intact. 
For higher values of disorder the dispersion breaks up as shown in Fig.~\ref{fig:disorder2} and Fig.~\ref{fig:disorder1}.  .

\section{Fermi surface}

It is amusing to note that from the perspective of quantum criticality, the ``normal" state is the superconducting state with a gap, $\Delta$ tied to the Fermi surface; the Fermi surface is obtained when the superconducting gap collapses: $\Delta(\lambda_1, \lambda_2, \lambda_3 \ldots)=0$, where $\lambda_1, \lambda_2, \lambda_3, \ldots$ are possible coupling constants.~\footnote{We are trying to illustrate a simple point regarding gaplessnes and  are aware of gapless superconductors as well as unique coherence properties of a superconductor in general}The surface spanned by more than two coupling constants could of course   be described as a quantum critical surface. In the case of BCS theory, only {\em one} coupling constant is relevant, in which case the critical surface converts to a QCP.
The essence of Kohn-Luttinger theory\cite{Kohn:1965} of superconductivity is that all Fermi systems, whether or not attractive or repulsive, has the ground state that is superconducting (as long as spatial inversion, time-reversal, or both symmetries are respected). The important point  in our present context  is the collapse of the gap, not the applicability of the BCS theory. 

The  Fermi surface viewed as a critical surface has been discussed in great detail by Volovik\cite{Volovik:2003} from the perspective of a topological object in the momentum space. The singularities of the Green function in $(p_0, {\bf p})$ frquency-momentum space is topologically equivalent to a quantized vortex in ($3+1$-dimensional) spacetime $({\bf r}, t)$. 

The important point about the Fermi surface is its lack of  scale;  it is infinitely sharp. As long as the modes do not couple, there are no phenomena that are out of the ordinary.  But when they do, we run into important effects, as in the Kondo problem.\cite{Wilson:1983}  However, some aspect of criticality of the Fermi surface can be gleaned from an elementary calculation\cite{Landau:1980} of  density-density correlation function of a Fermi system:

\begin{equation}
\left<\Delta n_1 \Delta n_2\right>  = \overline n \delta({\bf r}_2 - {\bf r}_1) +  \overline n \nu(r),
\end{equation}
where $\overline n$ is the average density, and $\Delta n(r)= (n(r) -\overline n)$,
\begin{equation}
\nu(r)=-\frac{g}{\overline n}\left |\int \frac{d^3p}{(2\pi\hbar)^3}\frac{e^{i {\bf p}\cdot {\bf r/}\hbar}}{e^{(\varepsilon_p - \mu)/T}+1}\right |^2,
\end{equation}  
$g=(2S+1)$; $S$ is the spin. As $T\to 0$, and $p_Fr/\hbar$ is large compared to $\varepsilon_F/T$ and also with respect  to unity:
\begin{equation}
\nu(r) \approx-\frac{3T^2}{\hbar v_F^2 p_F r^2} \exp \left(-\frac{2\pi  T r}{\hbar v_F}\right).
\end{equation}
Here $v_F$ is the Fermi velocity and $p_F$  is the Fermi momentum. The criticality of the Fermi surface is manifest in the  correlation length $\xi= \hbar v_F/2\pi  T$, which diverges as $T\to 0$. The corresponding time scale is     given by 
\begin{equation}
\tau = \frac{\xi}{v_F}= \frac{\hbar }{2\pi  T},
\end{equation}
which is an elementary example of what is known as planckian time discussed in the recent literature.~\cite{Zaanen:2019,Hartnoll:2022} The above result  is typical of a gapless  system, and is applicable to many physical systems; see, for example, the $(2+1)$-dimensional quantum Heisenberg model.~\cite{Chakravarty:1989} This Planckian time is not the momentum relaxation time in a Fermi liquid, $1/\tau_F\propto A (1/\hbar)T^2/\varepsilon_F$, as discussed in textbooks.\cite{Ashcroft:1976} $A$ is a dimensionless constant.

\section{Summary}
In this paper we have have explored an exactly solved model that exhibits three interesting quantum critical lines and two multicritical points. The centerpiece is the notion that the existence of quantum critical lines allow one to explore zero temperature quantum critical fluctuations without excessive fine tuning, as would be the case for a quantum critical point. The three lines have their unique characteristics. On one line criticality is unchanged and are located at $k=\pm \pi$ ($\nu=1, z=1$), in the other it is centered at $k=0$ ($\nu=1, z=1$),  and in the third the criticality is at incommensurate $k$-points. It is remarkable that the same model can exhibit such varied behavior. In addition there two multicritical points. One of which corresponds to nonrelativistic quadratic dispersion with a dynamical exponent $z=2$ and a critical exponent $\nu=1/2$. The specific heat on all critical lines except at the multicritical point $(d)$ in Fig.~\ref{fig:fig1} is linear in $T$ with differing slopes; at this multicritical point, the specific heat is proportional to $\sqrt{T}$.

The transition lines at $T=0$ are  topological in the sense that the number of Majorana zero modes at each end of the chain changes across the transition lines. Since all realistic physical examples must involve some degree of disorder, we explored its effects on the dispersion spectrum. It is quite fortunate that the real-time spectra can be calculated because the model is exactly solved. Typically it is difficult to calculate the real-time spectra. 

We believe that experimental realizations of the model can be found in which a free chain is all that is needed. Perhaps experimental techniques of NMR relaxation methods\cite{Kinross:2014} as well as terahertz spectroscopy could be employed.\cite{Armitage:2014,Armitage:2022} The artifact of periodic boundary condition is not necessary, simplifying the search for a physical model.

It is possible to extend our model by adding further neighbor interactions,  still maintaining its exact solvability, so as to discuss quantum critical surfaces in the parameter space. However, experimental realizations of such  models will be increasingly difficult to achieve. Finally, QCS that is only cursorily mentioned  here could be found in the language of gauge-gravity dual ideas.\cite{Zaanen:2015}

\begin{acknowledgments}
The authors would like to thank S. Raghu for useful comments. H.Y. was supported by Mani L. Bhaumik Institute for theoretical Physics at UCLA. S. C. was supported from the funds from the David S. Saxon Presidential Term Chair.
\end{acknowledgments}

%

\end{document}